\documentclass[trackchanges]{aastex631}

\usepackage{amsmath} 
\usepackage{subfigure}
\usepackage{bm}
\usepackage{float}

\begin{document}

\shortauthors{Shah et al.}

\shorttitle{Probing the maximum energy of FRBs}

\title{Probing the maximum energy of fast radio bursts using thousands of sources from the Second CHIME/FRB Catalog}

\newcommand{\TAU}{\affiliation{School of Physics and Astronomy, Tel Aviv University, Tel Aviv 69978, Israel}}

\newcommand{\SETI}{\affiliation{SETI Institute, Mountain View, CA, USA}}

\newcommand{\CALTECH}{\affiliation{Cahill Center for Astronomy and Astrophysics, MC 249-17 California Institute of Technology, Pasadena CA 91125, USA}}

\newcommand{\ICRAR}{\affiliation{International Centre for Radio Astronomy Research (ICRAR), Curtin University, Bentley WA 6102 Australia}}

\newcommand{\DRAO}{\affiliation{Dominion Radio Astrophysical Observatory, Herzberg Research Centre for Astronomy and Astrophysics, National Research Council Canada, PO Box 248, Penticton, BC V2A 6J9, Canada}}

\newcommand{\CSIRO}{\affiliation{CSIRO Space \& Astronomy, Parkes Observatory, P.O. Box 276, Parkes NSW 2870, Australia}}

\newcommand{\NRC}{\affiliation{NRC Herzberg Astronomy and Astrophysics, 5071 West Saanich Road, Victoria, BC V9E2E7, Canada}}

\newcommand{\CIERA}{\affiliation{Center for Interdisciplinary Exploration and Research in Astronomy (CIERA), Northwestern University, 1800 Sherman Avenue, Evanston, IL 60201, USA }}

\newcommand{\NU}{\affiliation{Department of Physics and Astronomy, Northwestern University, Evanston, IL 60208, USA}}

\newcommand{\Uch}
{\affiliation{Department of Astronomy and Astrophysics, University of Chicago, William Eckhart Research Center, 5640 South Ellis Avenue, Chicago, IL 60637, USA}}

\newcommand{\UCSC}{\affiliation{Department of Astronomy and Astrophysics, University of California Santa Cruz, 1156 High Street, Santa Cruz, CA 95064, USA}}

\newcommand{\IPMU}{\affiliation{Kavli Institute for the Physics and Mathematics of the Universe (Kavli IPMU), 5-1-5 Kashiwanoha, Kashiwa, 277-8583, Japan}}

\newcommand{\NAOJ}{\affiliation{Division of Science, National Astronomical Observatory of Japan, 2-21-1 Osawa, Mitaka, Tokyo 181-8588, Japan}}

\newcommand{\MU}{\affiliation{Department of Physics, McGill University, 3600 rue University, Montr\'eal, QC H3A 2T8, Canada}}

\newcommand{\Trottier}{\affiliation{Trottier Space Institute, McGill University, 3550 rue University, Montr\'eal, QC H3A 2A7, Canada}}

\newcommand{\CMU}{\affiliation{McWilliams Center for Cosmology \& Astrophysics, Department of Physics, Carnegie Mellon University, Pittsburgh, PA 15213, USA}}

\newcommand{\UVA}
{\affiliation{Anton Pannekoek Institute for Astronomy, University of Amsterdam, Science Park 904, 1098 XH Amsterdam, The Netherlands}}

\newcommand{\ASTRON}
{\affiliation{ASTRON, Netherlands Institute for Radio Astronomy, Oude Hoogeveensedijk 4, 7991 PD Dwingeloo, The Netherlands
}}

\newcommand{\MITK}
{\affiliation{MIT Kavli Institute for Astrophysics and Space Research, Massachusetts Institute of Technology, 77 Massachusetts Ave, Cambridge, MA 02139, USA}}

\newcommand{\MITP}
{\affiliation{Department of Physics, Massachusetts Institute of Technology, 77 Massachusetts Ave, Cambridge, MA 02139, USA}}

\newcommand{\CCAPS}{\affiliation{Cornell Center for Astrophysics and Planetary Science, Cornell University, Ithaca, NY 14853, USA}}

\newcommand{\DI}
{\affiliation{Dunlap Institute for Astronomy and Astrophysics, 50 St. George Street, University of Toronto, ON M5S 3H4, Canada}}

\newcommand{\DAA}
{\affiliation{David A. Dunlap Department of Astronomy and Astrophysics, 50 St. George Street, University of Toronto, ON M5S 3H4, Canada}}

\newcommand{\STSCI}
{\affiliation{Space Telescope Science Institute, 3700 San Martin Drive, Baltimore, MD 21218, USA}}

\newcommand{\WVUPHAS}
{\affiliation{Department of Physics and Astronomy, West Virginia University, PO Box 6315, Morgantown, WV 26506, USA }}

\newcommand{\WVUGWAC}
{\affiliation{Center for Gravitational Waves and Cosmology, West Virginia University, Chestnut Ridge Research Building, Morgantown, WV 26505, USA}}

\newcommand{\UCB}
{\affiliation{Department of Astronomy, University of California, Berkeley, CA 94720, United States}}

\newcommand{\YORK}
{\affiliation{Department of Physics and Astronomy, York University, 4700 Keele Street, Toronto, ON MJ3 1P3, Canada}}

\newcommand{\PI}
{\affiliation{Perimeter Institute of Theoretical Physics, 31 Caroline Street North, Waterloo, ON N2L 2Y5, Canada}}

\newcommand{\UBC}
{\affiliation{Department of Physics and Astronomy, University of British Columbia, 6224 Agricultural Road, Vancouver, BC V6T 1Z1 Canada}}

\newcommand{\UCHILE}
{\affiliation{Department of Electrical Engineering, Universidad de Chile, Av. Tupper 2007, Santiago 8370451, Chile}}

\author[0000-0002-4823-1946]{Vishwangi~Shah}
\MU
\Trottier
%\email{vishwangi.shah@mail.mcgill.ca}

\author[0000-0003-2317-1446]{Jason~W.~T.~Hessels}
\MU
\Trottier
\UVA
\ASTRON

\author[0000-0001-9345-0307]{Victoria~M.~Kaspi}
\MU 
\Trottier
\TAU

\author[0000-0002-4279-6946]{Kiyoshi~W.~Masui}
\MITK
\MITP

\author[0000-0002-4623-5329]{Mawson~W.~Sammons}
\MU
\Trottier

\author[0009-0005-5370-7653]{Daniel~Amouyal}
\MU
\Trottier

\author[0000-0002-1800-8233]{Charanjot~Brar}
\NRC

\author[0000-0002-2878-1502]{Shami~Chatterjee}
\CCAPS

\author[0000-0002-8376-1563]{Alice~P.~Curtin}
\MU
\Trottier

\author[0009-0002-1640-1776]{Hannah~Didehbani}
\MITP

\author[0000-0002-3382-9558]{B.~M.~Gaensler}
\UCSC
\DAA
\DI

\author[0009-0009-0938-1595]{Naman~Jain}
\MU
\Trottier

\author[0000-0003-3457-4670]{Ronniy~C.~Joseph}
\MU
\Trottier

\author[0009-0004-4176-0062]{Afrokk~Khan}
\MU
\Trottier

\author[0009-0008-6166-1095]{Bikash~Kharel}
\WVUPHAS
\WVUGWAC

\author[0000-0003-2116-3573]{Adam~E.~Lanman}
\MITK
\MITP

\author[0000-0003-2111-3437]{Kyle~McGregor}
\MU
\Trottier

\author[0000-0001-7348-6900]{Ryan~Mckinven}
\MU
\Trottier

\author[0000-0002-0940-6563]{Mason~Ng}
\MU
\Trottier

\author[0000-0003-0510-0740]{Kenzie~Nimmo}
\CIERA

\author[0000-0002-8897-1973]{Ayush~Pandhi}
\MU
\Trottier

\author[0000-0002-8912-0732]{Aaron~B.~Pearlman}
\altaffiliation{NASA Hubble Fellow.}
\MU
\Trottier
\MITK
\MITP
% \email{aaron.b.pearlman@mit.edu}

\author[0000-0002-3430-7671]{Alexander~W.~Pollak}
\SETI

\author[0000-0002-7374-7119]{Paul~Scholz}
\YORK

\author[0000-0002-6823-2073]{Kaitlyn~Shin}
\CALTECH

\author[0000-0003-2631-6217]{Seth~R.~Siegel}
\MU
\Trottier
\PI

\author[0000-0002-2088-3125]{Kendrick~Smith}
\PI

\author[0009-0001-3334-9482]{Michele~Woodland}
\UCSC
\correspondingauthor{Vishwangi Shah}
\email{vishwangi.shah@mail.mcgill.ca}

\begin{abstract}

Quantifying the maximum energy of fast radio bursts (FRBs) can provide stringent constraints on their emission mechanisms and progenitor models. However, the most energetic bursts are rare, requiring a large sample of FRBs to detect them. In this work, we use the largest available such sample, 2,998 one-off FRBs from the Second CHIME/FRB Catalog, to obtain a lower limit on the maximum energy ($E^{\mathrm{max}}_{\mathrm{iso}}$) of FRBs, assuming isotropic energy distribution from FRB sources. In the absence of known redshifts ($z$) for most sources, we present a framework that uses the dispersion measures (DMs) and fluences of these FRBs, together with the probability distribution of $z$ given DM, to derive the lower limit on $E^{\mathrm{max}}_{\mathrm{iso}}$. We generate simulated FRB samples assuming different parameter values for a log-normal $\mathrm{DM}_{\mathrm{host}}$ distribution and a Schechter function form of the FRB energy function to estimate how many outliers -- FRBs with large DM contributions from the host galaxy or intervening galaxy halos -- could artificially inflate this limit. After accounting for outliers, the lower limit on $E^{\mathrm{max}}_{\mathrm{iso}}$ from Catalog 2 FRBs ranges between $1.2\times10^{41}$ and $1.9\times10^{42}$\,erg, with best estimate $1.2\times10^{42}$\,erg. This limit is consistent with those derived from much smaller FRB samples. Moreover, inferred energies of hundreds of FRBs appear collectively limited around $\sim10^{42}$\,erg, suggesting a physical limit on the energy reservoir of FRB sources. The corresponding isotropic-equivalent FRB source energy is consistent with the total energy available in a magnetar's external dipole magnetic field, supporting magnetars as FRB progenitors. 

\end{abstract}

\keywords{Radio bursts (1339) --- Radio transient sources (2008) --- Magnetars (992) --- High energy astrophysics(739)}

\section{Introduction} 

  Fast radio bursts (FRBs) are extremely energetic radio transients, having isotropic equivalent energies ranging over $10^{37}-10^{42}$\,erg, packed into millisecond time durations, making them detectable at cosmological distances of millions to billions of parsecs \citep{Petroff_2022}. First detected in 2007 \citep{Lorimer_2007}, thousands of FRBs are now reported \citep{chime_cat2}; however, their origins remain a mystery. About $\sim3\%$ of FRBs have been observed to repeat \citep{cat1, RN3}, ruling out cataclysmic origins for at least this class of FRBs. The clues for FRB progenitors and emission mechanisms can be found in their burst properties as well as the kind of galaxies and local environments in which they occur. Their millisecond duration points towards compact object progenitors, and microsecond-duration bursts and sub-bursts observed from some FRB sources \citep{Nimmo_M81_microshots, Snelders_2023} as well as a neutron-star like polarization position angle swing in at least one source \citep{Mckinven_PA} further provide evidence that the bursts are generated in the magnetospheres of neutron stars. The high brightness temperatures ($T_B > 10^{32}$\,K) of FRBs imply a coherent emission process, such as reconnection in the neutron star magnetosphere \citep{Lyutikov_Popov_2020, Lyutikov_2021}, or magnetar flares colliding with the surrounding medium \citep{Lyubarsky_maser, Metzger_2019}. The preferential occurrence of FRBs in star-forming galaxies further supports magnetars formed via core-collapse supernovae of massive stars as progenitors \citep{heintz_2020, Bhandari_2022, Gordon_2023, Mohit_host_galaxy, Kritti_DSA_host_gal}. However, some FRBs localized to quiescent environments suggest alternative formation channels for FRB progenitors \citep{Kirsten_M81, Kritti_DSA_host_gal, Shah_R155, Tarraneh_R155}. 

  The maximum isotropic energy ($E^{\mathrm{max}}_{\mathrm{iso}}$) of FRBs can provide an effective way to examine the viability of the various FRB emission mechanisms and progenitor models. FRBs could be beamed, but their beaming fraction is unknown, so isotropic emission is typically assumed when estimating energy. The most energetic FRBs yet detected, having $E_{\mathrm{iso}} \sim10^{42}$\,erg \citep{Ravi_2019, Ryder_2023} are reaching the limits of what a magnetar could produce from external dipolar magnetic fields, given the standard assumptions for radio efficiencies \citep[$\eta_{\mathrm{radio}} \lesssim 10^{-4}$,][]{Bochenek_SGR} and surface dipolar field strength ($B_P = 10^{15}$\,G). Bursts more energetic than this could imply extreme values for these parameters, energy released from stronger multi-pole components of the magnetar magnetic field, upper limits on the beaming fraction of FRBs, or a different progenitor altogether. However, the energy range of detected FRBs is limited by the sensitivity of radio telescopes on one end, and by the low rates of highly energetic bursts on the other end \citep{James_2022, Shin_FRB_pop, Kirsten_2024, Omar_2024}. The rare nature of the most energetic bursts necessitates a large sample of FRBs to be able to detect them. The Second CHIME/FRB Catalog (hereafter referred to as Catalog 2) comprises 4,539 FRBs originating from 3,641 unique sources \citep{chime_cat2}, and provides an opportunity to probe unusually energetic FRBs. 
  
  In this work, we estimate a lower limit on $E^{\mathrm{max}}_{\mathrm{iso}}$ using Catalog 2 FRBs. The dataset used for this purpose is discussed in Section~\ref{data}; the methods used to estimate and verify our results are described in Section~\ref{Methods}. The lower limit on $E^{\mathrm{max}}_{\mathrm{iso}}$ value that we derive is presented in Section~\ref{results} and the resulting implications for the progenitors of FRBs are discussed in Section~\ref{Discussion}. 

\section{Dataset}\label{data}

The Canadian Hydrogen Intensity Mapping Experiment (CHIME) is a transit radio telescope located in Penticton, British Columbia and operating in the $400-800$\,MHz frequency band \citep{chime_overview}. It consists of four 100\,m $\times$ 20\,m cylindrical paraboloidal reflectors that provide a $\gtrsim$ 200\,deg$^2$ instantaneous field of view (FOV). CHIME's FRB search engine makes use of this large FOV, its powerful correlator, and wide bandwidth to search the entire Northern sky every day for FRBs \citep{chime/frb_overview}. This makes CHIME/FRB the leader in FRB discovery rate, observing $\sim3$ FRBs per day. 

Catalog 2 \citep{chime_cat2} reports on 4,539 bursts, including 981 bursts from 83 known repeating sources, detected by CHIME/FRB between 2018 July and 2023 September. It also characterizes the properties of these bursts, including DMs, temporal widths, scattering times, fluences, etc., and presents a $O(10')$ sky localization for each source. The large localization uncertainty prevents host galaxy association and thus redshift ($z$) determination for most of these FRBs. The reported fluences are lower limits, as in the absence of accurate sky location, the bursts are assumed to be detected at the location of the primary beam's highest sensitivity in the FRB's declination arc on the sky \citep{bridget_flux}. 

In this work, we use the DMs and fluence lower limits of one-off FRBs presented in Catalog 2. We decided to exclude repeating FRBs because they are known to often have distinct burst morphologies compared to apparently one-off FRBs \citep{Ziggy_repeater_morphology, Curtin_2025}, possibly resulting in different detection biases between the two sub-samples. Our sample also excludes FRBs that have extragalactic DM less than 100\,pc\,cm$^{-3}$. Additionally, we remove FRBs that do not have reported fluence measurements due to calibration issues or RFI contamination, and FRBs detected during periods of non-nominal system sensitivity, those discovered by citizen scientists, and those detected in the far sidelobes of the primary beam \citep{chime_cat2}. This leaves 2,998 unique FRB sources having the required DM and fluence information, which we then combine with the probability distribution function (PDF) of $z$ given DM, $P(z \mid \mathrm{DM})$, to estimate a lower limit on $E^{\mathrm{max}}_{\mathrm{iso}}$, as described in Section~\ref{Methods}.  

\section{Methods}\label{Methods}
In Section~\ref{framework}, we present our framework for obtaining a lower limit on $E^{\mathrm{max}}_{\mathrm{iso}}$ using the DMs and fluences of a sample of FRBs. To test the performance of this method, we apply it to simulated FRB samples and compare the obtained lower limit on $E^{\mathrm{max}}_{\mathrm{iso}}$ to the true $E^{\mathrm{max}}_{\mathrm{iso}}$ in the sample. The simulated samples are generated using the steps and assumptions described in Section~\ref{simulation}. In Section~\ref{outliers}, we develop a method to use the simulated samples to estimate the number of outlier FRBs that can falsely inflate the lower limit on $E^{\mathrm{max}}_{\mathrm{iso}}$ due to excess DM from the host or intervening halos. We obtain the lower limit on $E^{\mathrm{max}}_{\mathrm{iso}}$ from Catalog 2 FRBs after accounting for the outliers. 

\subsection{Framework to Obtain Lower Limit on $E^{\mathrm{max}}_{\mathrm{iso}}$} \label{framework}

The observed DM of an FRB can be expressed as the sum of various components, 
\begin{equation}\label{total_DM}
    \mathrm{DM}(z) = \mathrm{DM}_{\mathrm{MW, ISM}} + \mathrm{DM}_{\mathrm{MW, halo}} + \mathrm{DM}_{\mathrm{cosmic}}(z) + \frac{\mathrm{DM}_{\mathrm{host}}}{1+z}.
\end{equation}
$\mathrm{DM}_{\mathrm{MW, ISM}}$ and $\mathrm{DM}_{\mathrm{MW, halo}}$ are the contributions from the interstellar medium (ISM) and halo of the Milky Way (MW), respectively. $\mathrm{DM}_{\mathrm{cosmic}}$ is from the ionized gas in the intergalactic medium (IGM) and any intervening galaxy halos along the line of sight. It depends on the path length through the IGM and thus is a function of  redshift ($z$). $\mathrm{DM}_{\mathrm{host}}$ is the DM contributed by the host galaxy of the FRB, which is converted to the observer's frame by the $1+z$ factor. 

To calculate the energies of FRBs, their redshifts are required. In the absence of known redshifts, we can use the dependence of $\mathrm{DM}$ on $z$ to obtain $P(z \mid \mathrm{DM})$, and translate that to a PDF of $E_{\mathrm{iso}}$ given DM, $P( E_{\mathrm{iso}} \mid \mathrm{DM})$, using the following steps:
\begin{enumerate}
    \item Using the Bayes' theorem:
    \begin{equation}
        P(z \mid \mathrm{DM}) \propto P(\mathrm{DM} \mid z)\times P(z).
    \end{equation}
    For a grid of $z$ ranging between 0.01 and 6, we assume a flat, uninformative prior for $P(z)$. \cite{Shin_FRB_pop} presents a $P(z \mid \mathrm{DM})$ for FRBs detected by CHIME/FRB, taking into account telescope-dependent selection biases. However, this $P(z \mid \mathrm{DM})$ includes an assumed luminosity function for FRBs and thus cannot be used to constrain the maximum energy of FRBs. We thus proceed with a telescope-independent $P(z \mid \mathrm{DM})$ and develop a method to account for selection effects post-facto in Sections~\ref{simulation}, \ref{outliers}, and \ref{results}.
    \item For $P(\mathrm{DM} \mid z)$, we marginalize over $\mathrm{DM}_{\mathrm{host}}$ and $\mathrm{DM}_{\mathrm{MW}}$ to obtain
    \begin{equation}
P(\mathrm{DM}|z) \propto \int \int d\mathrm{DM}_{\mathrm{MW}}\,d\mathrm{DM}_{\mathrm{host}}\,P( \mathrm{DM}_{\mathrm{MW}})\,P\left( \mathrm{DM}_{\mathrm{cosmic}} = \mathrm{DM} - \mathrm{DM}_{\mathrm{MW}} - \frac{\mathrm{DM}_{\mathrm{host}}}{1+z} \mid z\right)\,P( \mathrm{DM}_{\mathrm{host}})
    \end{equation} 
    where $\mathrm{DM}_{\mathrm{MW}}$ is the sum of $\mathrm{DM}_{\mathrm{MW, ISM}}$ and $\mathrm{DM}_{\mathrm{MW, halo}}$. We present a more detailed derivation of this form of $P(\mathrm{DM} \mid z)$ in Appendix~\ref{sec:P_DM_z}.  
    \item We use a parameterization of $P( \mathrm{DM}_{\mathrm{cosmic}} \mid z)$ \citep{McQuinn_2014, Macquart_2020} that takes into account filaments and intervening halos to give a probabilistic distribution of ionized gas in the Universe about a mean value:
    \begin{equation}
        P( \mathrm{DM}_{\mathrm{cosmic}} \mid z) = \langle \mathrm{DM}_{\mathrm{cosmic}} \rangle P(\Delta_{\mathrm{DM}}).
    \end{equation}
    $\langle \mathrm{DM}_{\mathrm{cosmic}} \rangle$ is the mean cosmological DM as a function of $z$, which assuming a flat Universe with matter and dark energy, can be written as
    \begin{equation}
        \langle \mathrm{DM}_{\mathrm{cosmic}} \rangle (z) = \int_0^z \frac{c\, \bar{n}_e(z')\, dz'}{H_0 (1 + z')^2 \sqrt{\Omega_m (1 + z')^3 + \Omega_\Lambda}}
    \end{equation}
    where $\bar{n}_e(z')$ is the mean density of free electrons, and we assume Planck cosmology with  $H_0 = 67.4$\,km\,s$^{-1}$, matter density $\Omega_m = 0.315$, and dark energy density $\Omega_\Lambda = 0.685$ \citep{Planck_2020}.  
    $\Delta_{\mathrm{DM}}$ is the deviation from the mean caused by intersections with galactic halos along the line of sight, such that $P(\Delta_{\mathrm{DM}})$ can be written as:
    \begin{equation}
        P(\Delta_{\mathrm{DM}})(z) = A \Delta_{\mathrm{DM}}^{-\beta} \exp\left[ -\frac{ \left( \Delta_{\mathrm{DM}}^{-\alpha} - C_0 \right)^2 }{ 2 \alpha^2 \sigma_{\mathrm{DM}}(z)^2 } \right]
    \end{equation}
    where we use the model parameter values from \cite{James_2022}. $\sigma_{\mathrm{DM}}(z) = Fz^{-0.5}$, where the feedback parameter $F$ quantifies the strength with which galaxies expel baryons. Our implementation of $P( \mathrm{DM}_{\mathrm{cosmic}} \mid z)$ is based on the \verb|grid_P_DMcosmic_z| function in the public \verb|FRB|\footnote{\url{https://github.com/FRBs/FRB/blob/main/frb/dm/prob_dmz.py}} GitHub repository. 
    \item We assume $\mathrm{DM}_{\mathrm{host}}$ values between 1 and 1,000\,pc\,cm$^{-3}$ with $P( \mathrm{DM}_{\mathrm{host}})$ given by a log-normal distribution of the form
    \begin{equation}\label{DM_host_pdf}
        P(\mathrm{DM}_{\mathrm{host}}) = \frac{1}{\mathrm{DM}_{\mathrm{host}}}\frac{1}{\sigma_{\mathrm{host}}\sqrt{2\pi}}e^{-\frac{\left(\mathrm{log}\mathrm{DM}_{\mathrm{host}} - \mu_{\mathrm{host}}\right)^2}{2\sigma_{\mathrm{host}}^2}}
    \end{equation}
    where $\mu_{\mathrm{host}} = 4.44$ and $\sigma_{\mathrm{host}} = 0.9$ in natural log space are the best-fit values from \cite{Shin_FRB_pop}. 
    \item We take $P( \mathrm{DM}_{\mathrm{MW}})$ to be the convolution of $P( \mathrm{DM}_{\mathrm{MW, ISM}})$ and $P( \mathrm{DM}_{\mathrm{MW, halo}})$, as implemented by \cite{Cordes_2022}. A $\mathrm{DM}_{\mathrm{MW, ISM}}$ estimate is obtained for each FRB using the NE2001 model for DM in the Milky Way ISM \citep{NE2001} and the Galactic coordinates of the FRB. $P( \mathrm{DM}_{\mathrm{MW, disk}})$ is a flat distribution with a 40\% spread centered on the NE2001 value. $P( \mathrm{DM}_{\mathrm{MW, halo}})$ is a flat distribution from 25\,pc\,cm$^{-3}$ to 80\,pc\,cm$^{-3}$, incorporating the various estimates for $\mathrm{DM}_{\mathrm{MW, halo}}$ \citep{DM_halo_prochaska_zheng, DM_halo_yamasaki_totani, DM_halo_cook, Shin_FRB_pop}.
    \item For each $z$ in $P(z \mid \mathrm{DM})$, the corresponding energy can be calculated as :
\begin{equation}\label{E}
    E_\text{iso}(z) = \frac{4\pi\text{D}_{\text{L}}^2}{(1+z)^{2+\alpha}}F_{\nu}\Delta\nu
\end{equation}
where $\text{D}_{\text{L}}$ is the luminosity distance, $F_{\nu}$ is the fluence, $\Delta\nu$ is the intrinsic bandwidth, and $\alpha$ is the spectral-index of the FRB. The FRB signal is integrated in the band-averaged time series to obtain the fluence, assuming $\alpha = 0$. We carry this assumption forward in the energy calculation as well. Additionally, to be consistent with other works \citep{James_2022, Shin_FRB_pop}, we assume an intrinsic bandwidth of 1\,GHz for FRB sources. Using change of variables for $P(z \mid \mathrm{DM})$, we obtain
\begin{equation}\label{P_E_DM}
    \begin{split}
    P(E_\text{iso} \mid \mathrm{DM}) &= P(z \mid \mathrm{DM}) \left| \frac{dz}{dE_\text{iso}}\right| \\
    &= P(z \mid \mathrm{DM})\frac{\text{D}_{\text{L}}H_0\sqrt{\Omega_m(1+z)^3 + \Omega_{\Lambda}}}{2E_\text{iso}(1+z)c}
    \end{split}
\end{equation}
where $c = 3\times10^8$\,m\,s$^{-1}$ is the speed of light. A detailed derivation for this form of $P(E_\text{iso} \mid \mathrm{DM})$ is presented in Appendix~\ref{sec:P_E_DM}. 
    \end{enumerate}

A lower limit on $E^{\mathrm{max}}_{\mathrm{iso}}$ is then estimated jointly from $P(E_\text{iso} \mid \mathrm{DM})$ for each FRB in the sample. To do so, we evaluate the probability that at least one FRB in the sample has an energy above a trial energy ($E_{\text{trial}}$) as:
\begin{equation}\label{E_max}
P(E_\text{iso} > E_{\text{trial}}) = 1 - \prod_{i=1}^{N}P(E_\text{iso} < E_{\text{trial}} \mid \text{DM})_{\text{FRB}_i}
\end{equation}
where $P(E_\text{iso} < E_{\text{trial}} \mid \text{DM})_{\text{FRB}_i}$ is the cumulative distribution function (CDF) associated with $P(E_\text{iso} \mid \mathrm{DM})$ for FRB $i$ and $N$ is the number of FRBs in the sample. For a given sample of bursts, $E_{\text{trial}}$ ranges between the minimum possible energy, obtained using the minimum FRB fluence in the sample and $z=0.01$, and the maximum possible energy, obtained using the maximum FRB fluence in the sample and $z=6$. $P(E_\text{iso} > E_{\text{trial}})$ is calculated for each $E_{\text{trial}}$. The maximum $E_{\text{trial}}$ value with $P(E_\text{iso} > E_{\text{trial}}) > 0.99$ is considered to be our estimate for the lower limit on $E^{\mathrm{max}}_{\mathrm{iso}}$. This a `lower limit' because, according to the method described above, there is a $\gtrsim99\%$ chance that there is at least 1 FRB in the sample with energy higher than this value. 

\subsection{Simulations to Verify the Framework}\label{simulation}
To verify the robustness of the framework described in Section~\ref{framework}, we apply it to simulated FRB samples and validate the obtained lower limit on $E^{\mathrm{max}}_{\mathrm{iso}}$ against the true $E^{\mathrm{max}}_{\mathrm{iso}}$ in the sample. The simulated FRB samples are generated in the following way:
\begin{enumerate} 
    \item We assume a Schechter function for the FRB energy distribution, which is a power-law with index $\gamma$ and an exponential energy cut-off at $E^{\mathrm{cut-off}}_{\mathrm{iso}}$ at the high-energy end.  We model $\text{DM}_{\text{host}}$ as a log-normal distribution with mean $\mu_\mathrm{host}$ and standard deviation $\sigma_\mathrm{host}$, as in Equation~\ref{DM_host_pdf}.
    \item We generate simulated samples using combinations of the 95\% confidence interval values for $E^{\mathrm{cut-off}}_{\mathrm{iso}}$, $\gamma$, $\mu_{\mathrm{host}}$, and $\sigma_{\mathrm{host}}$, obtained from the posterior probability distributions from \cite{Shin_FRB_pop}. These parameter values were chosen so that our framework can be tested for simulations covering the extreme case scenarios for the $\mathrm{DM}_{\mathrm{host}}$ distribution and FRB energy function. We thus obtain 16 different simulated samples as specified in Table \ref{tab:Emax_simulated}. Additionally, we obtain a simulated sample for the best-fit parameter values from \cite{Shin_FRB_pop}. 
    \item  For each simulated FRB, we draw $z$ from a grid ranging between 0.01 and 6 according to the redshift probability distribution function expected for a given FRB population observed by CHIME/FRB. To calculate the expected redshift distribution, we use the zDM \citep{James_2022} package, assuming population and survey parameters as specified in \cite{Shin_FRB_pop}, but varying $E^{\mathrm{cutoff}}_{\mathrm{iso}}$, $\gamma$, $\mu_\mathrm{host}$ and $\sigma_\mathrm{host}$ to the combinations in Table \ref{tab:Emax_simulated}.
    \item For each FRB in a simulated sample, we draw an energy value from the corresponding Schechter function such that the obtained fluence is greater than 3.5\,Jy\,ms, the median 95\% fluence threshold of Catalog 2 FRBs \citep{chime_cat2}. 
     \item We take each of $\text{DM}_{\text{MW, ISM}}$ and $\text{DM}_{\text{MW, halo}}$ to be 50\,pc\,cm$^{-3}$.
    \item We draw $\mathrm{DM}_{\mathrm{cosmic}}$ from a grid ranging between 1\,pc\,cm$^{-3}$ and 5,000\,pc\,cm$^{-3}$ where the probability of a value being chosen is given by the $P( \mathrm{DM}_{\mathrm{cosmic}} \mid z)$ described in Section~\ref{framework}. 
    \item For a grid ranging between 1\,pc\,cm$^{-3}$ and 1,000\,pc\,cm$^{-3}$, we draw $\text{DM}_{\text{host}}$ from the log-normal distribution corresponding to the particular simulated sample. The total DM is then given by Equation~\ref{total_DM}. 
    \item For each simulated sample, we only select simulated FRBs that have extragalactic DM greater than 100 and less than or equal to the maximum extragalactic DM of 3,167\,pc\,cm$^{-3}$ in the subsample of Catalog 2 FRBs used in this work. This ensures that the simulated FRBs cover the same range of extragalactic DMs as the Catalog 2 FRBs. We thus obtain simulated samples of size $N = 2,998$ FRBs each, equal to the size of the real FRB sample we have.  
\end{enumerate}
An example of the framework applied to a simulated FRB sample is shown in Figure~\ref{fig:Emaxll}. The $P(E_\text{iso} > E_{\text{trial}})$ (grey curve) is equal to 1 upto a certain $E_{\text{trial}}$ because for these $E_{\text{trial}}$ values there is a $100\%$ chance that at least 1 FRB in the sample has a higher energy. The maximum $E_{\text{trial}}$ where  $P(E_\text{iso} > E_{\text{trial}}) > 0.99$ is our estimate for the lower limit on $E^{\mathrm{max}}_{\mathrm{iso}}$ (blue dashed line), which is also where $P(E_\text{iso} > E_{\text{trial}})$ starts falling rapidly before dropping to 0. The lower limit on $E^{\mathrm{max}}_{\mathrm{iso}}$ is slightly higher than the true $E^{\mathrm{max}}_{\mathrm{iso}}$ (black dashed line) in the sample. This is because, in some cases, it is possible for an FRB to have large DM contributions from the host galaxy or intervening galaxy halos such that the $P(z \mid \mathrm{DM})$ favors a larger $z$ than the true $z$ of the FRB. These FRBs can falsely seem more energetic than they truly are. In Section~\ref{outliers}, we develop a method to use the simulated samples to estimate the number of such outliers which we then remove from the Catalog 2 sample to obtain a realistic value for the lower limit on $E^{\mathrm{max}}_{\mathrm{iso}}$.

\subsection{Removing Outliers}\label{outliers}
To constrain the maximum energy of FRBs, we must exclude FRBs for which $P(E_\text{iso} \mid \mathrm{DM})$ favors a higher energy than their true energy due to larger than expected $\mathrm{DM}_{\mathrm{host}}$ and/or $\mathrm{DM}_{\mathrm{cosmic}}$. In the absence of known $z$ for most Catalog 2 FRBs, it is not possible to identify which FRBs are affected by this issue. Since we don't know which FRBs need to be removed from the Catalog 2 sample, we use the simulated samples to inform the number of FRBs that need to be removed. 

\begin{figure}[t]
    \centering
    \subfigure[]{\includegraphics[width=0.32\textwidth]{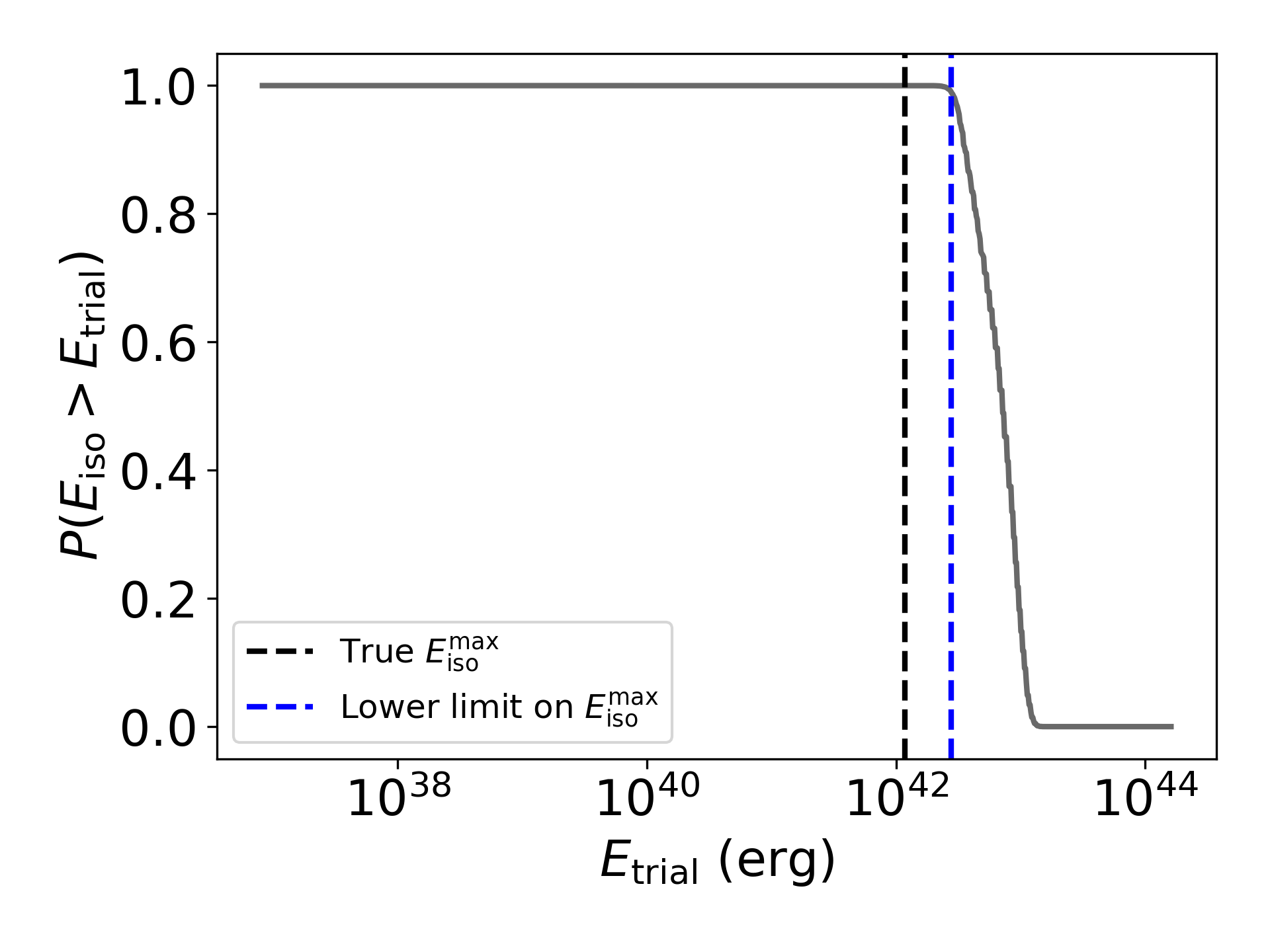}\label{fig:Emaxll}} 
    \subfigure[]{\includegraphics[width=0.32\textwidth]{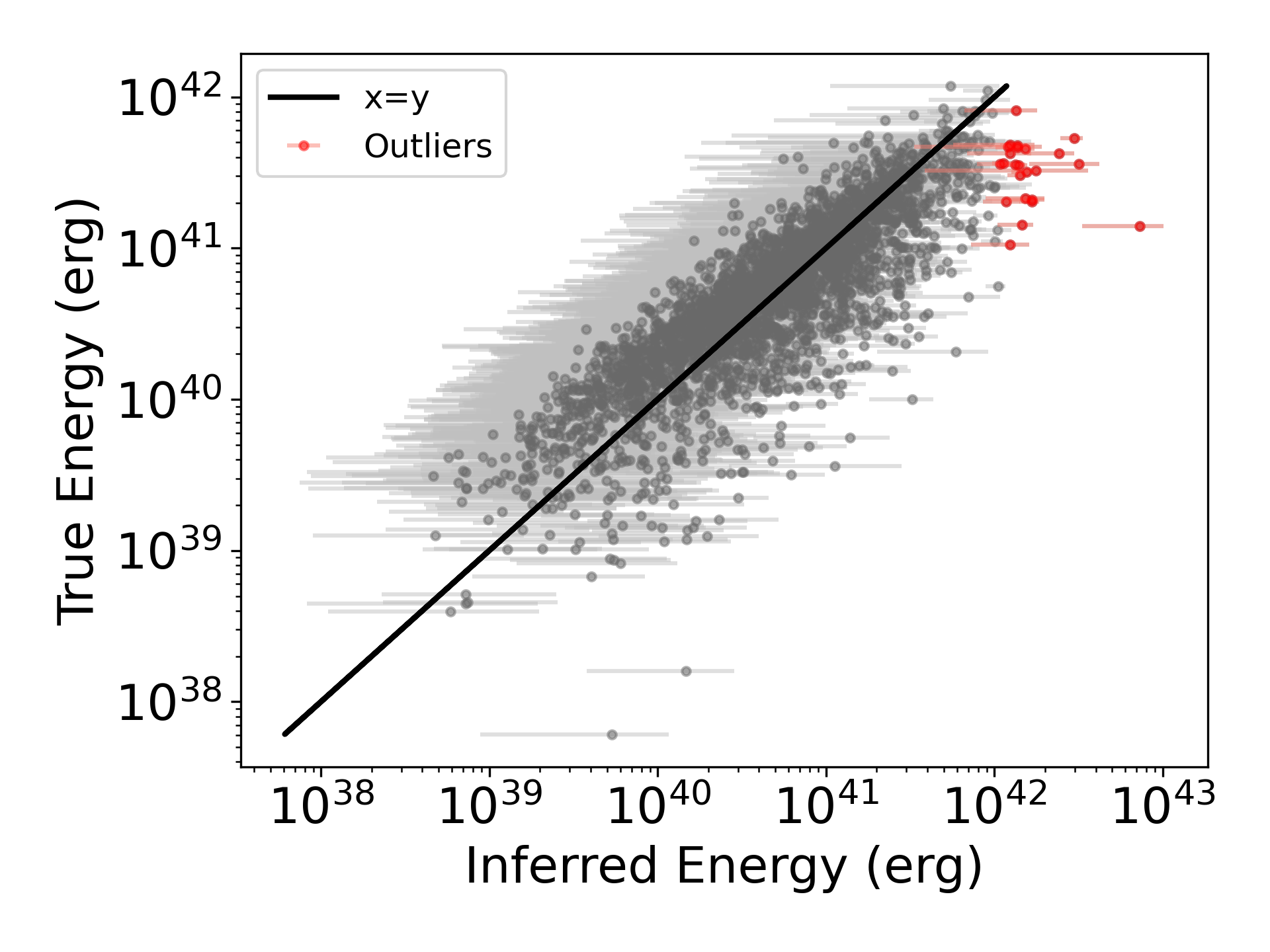}\label{fig:outliers}} 
    \subfigure[]{\includegraphics[width=0.32\textwidth]{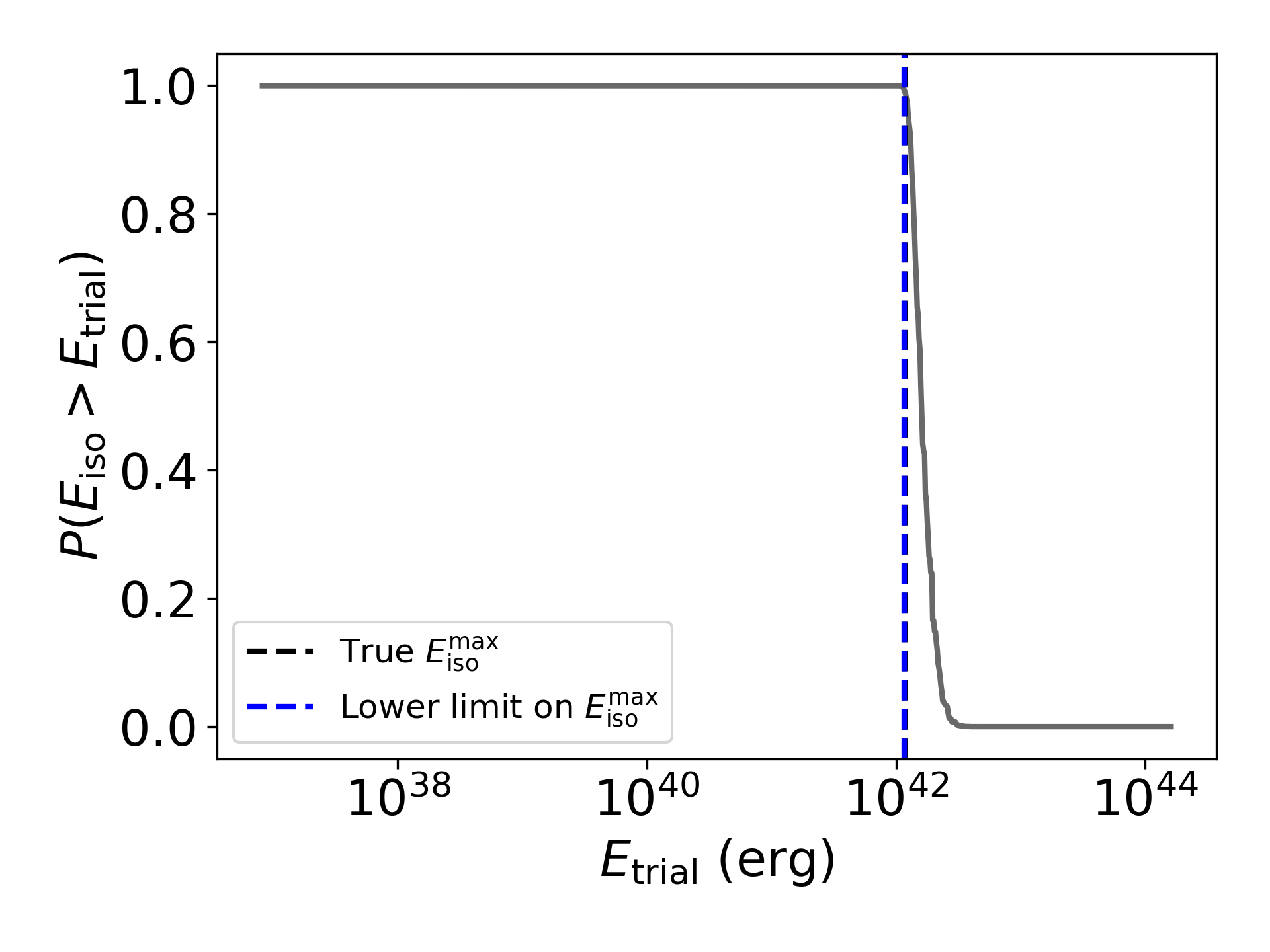}\label{fig:Emaxll_out}}
    \caption{For the simulated FRB sample having the best-fit values of $\mu_{\mathrm{host}}$, $\sigma_{\mathrm{host}}$, $E^{\mathrm{cut-off}}_{\mathrm{iso}}$, and $\gamma$ from \cite{Shin_FRB_pop}, Panel~\ref{fig:Emaxll} shows $P(E_{\text{iso}} > E_{\text{trial}})$ (gray line) before removing the outliers. The lower limit on $E^{\mathrm{max}}_{\mathrm{iso}}$ (blue dashed line) is slightly higher than the true $E^{\mathrm{max}}_{\mathrm{iso}}$ (black dashed line). Panel~\ref{fig:outliers} shows the median inferred energies with 68\% confidence interval from $P(E_\text{iso} \mid \mathrm{DM})$ for the simulated FRBs plotted against their true energies. The black line shows where the inferred and true energies are equal. The highest inferred FRBs which are considered outliers are marked in red. Panel~\ref{fig:Emaxll_out} shows $P(E_{\text{iso}} > E_{\text{trial}})$ (gray line) after removing the outliers. The lower limit on $E^{\mathrm{max}}_{\mathrm{iso}}$ (blue dashed line) in this case is consistent and overlaps with the true $E^{\mathrm{max}}_{\mathrm{iso}}$ (black dashed line). This describes our method of obtaining a realistic value for the lower limit on $E^{\mathrm{max}}_{\mathrm{iso}}$ for a simulated sample after removing a sub-sample of highest inferred energy bursts as outliers.\label{fig:fig_1}}
\end{figure}

To do so, we find the median energies inferred from $P(E_\text{iso} \mid \mathrm{DM})$ for each FRB in a simulated sample. The FRBs which have the highest inferred energies affect the lower limit on $E^{\mathrm{max}}_{\mathrm{iso}}$ estimate the most. Thus, we remove the highest inferred energy FRBs from the simulated samples until the lower limit on $E^{\mathrm{max}}_{\mathrm{iso}}$ obtained from the remaining FRBs is less than or equal to the true $E^{\mathrm{max}}_{\mathrm{iso}}$ in the sample. The FRBs that are removed are considered outliers. An example of this process is shown in Figure~\ref{fig:fig_1} for a simulated sample. The lower limit on $E^{\mathrm{max}}_{\mathrm{iso}}$ (blue dashed line) obtained in Figure~\ref{fig:Emaxll} before removing the outliers is greater than the true $E^{\mathrm{max}}_{\mathrm{iso}}$ (black dashed line). Figure~\ref{fig:outliers} shows the median inferred energy with 68\% confidence interval plotted against the true energies of the FRBs in the simulated sample. The black line shows where the true and inferred energies are equal. The FRBs with the highest inferred energies that are considered outliers (red points) are all below the black line, having inferred energies greater than the true energies. The lower limit on $E^{\mathrm{max}}_{\mathrm{iso}}$ (blue dashed line) obtained in Figure~\ref{fig:Emaxll_out} after removing the outliers is consistent and overlaps with the true $E^{\mathrm{max}}_{\mathrm{iso}}$ (black dashed line) in the simulated sample. We similarly obtain the number of outliers to be removed for all of our simulated samples. For each case, we confirm that the majority of the highest inferred energy FRBs that are considered outliers have inferred energies greater than their true energies, that is, they are truly contaminants. Thus, for different assumptions of $\mathrm{DM}_{\mathrm{host}}$ distribution and energy function of FRBs, this method allows us to estimate the number of contaminants with the highest inferred energies that need to be removed in order to obtain a realistic value for the lower limit on $E^{\mathrm{max}}_{\mathrm{iso}}$ from Catalog 2 FRBs.

\section{Results}\label{results} 
\begin{figure}[t]
\centering
\includegraphics[width=0.7\textwidth]{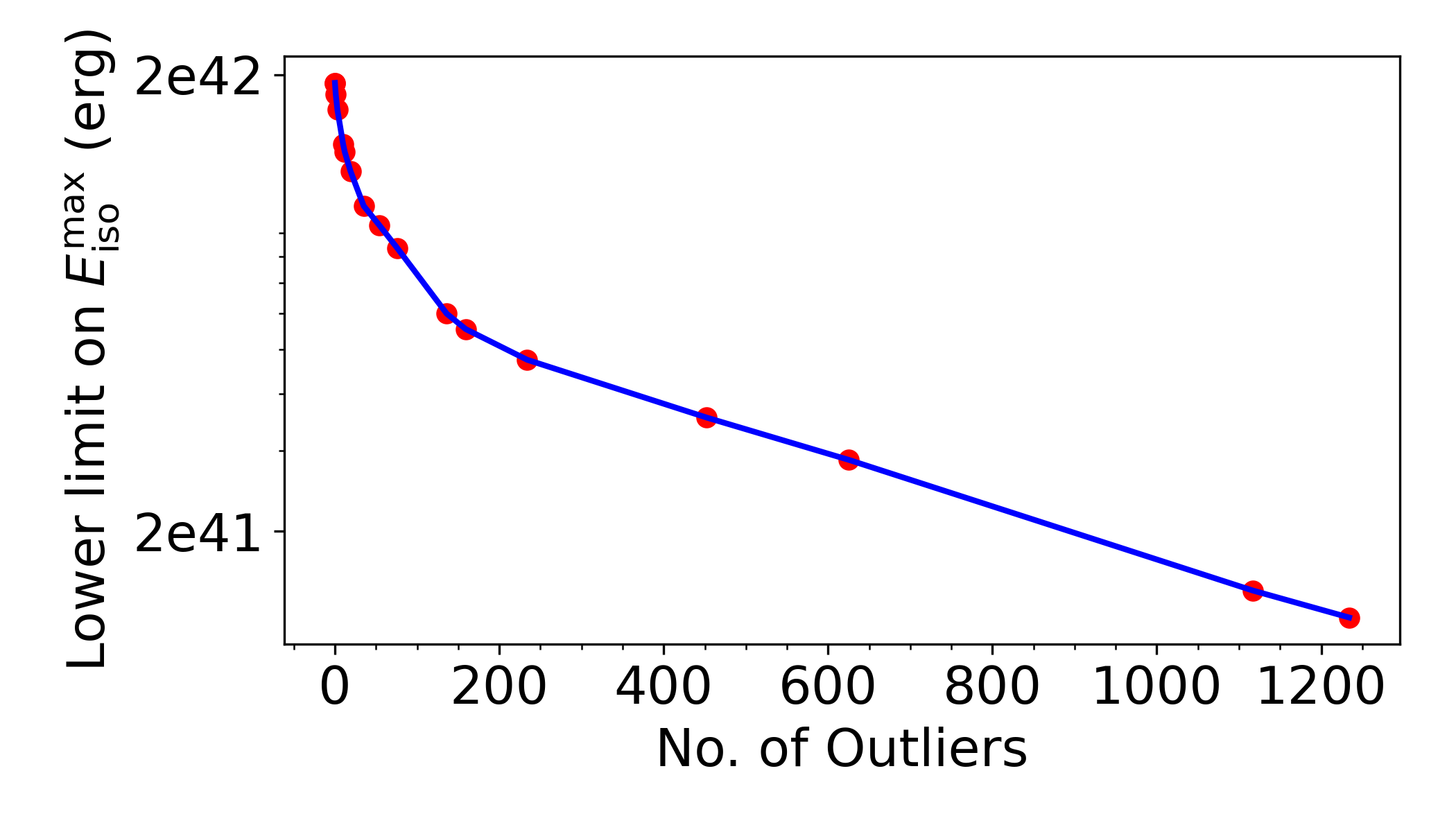}
\caption{The lower limit on $E^{\mathrm{max}}_{\mathrm{iso}}$ from Catalog 2 FRBs as a function of the number of highest inferred energy FRBs that are removed from the sample as outliers. The red points mark the lower limit value for the number of outliers obtained from the 17 simulated FRB samples. \label{fig:ll_vs_out}}
\end{figure}

In Table~\ref{tab:Emax_simulated} we present the lower limit on $E^{\mathrm{max}}_{\mathrm{iso}}$ from Catalog 2 FRBs after removing the highest inferred energy data points as outliers, where the number of outliers is informed from the 17 simulated samples with different $\text{DM}_{\text{host}}$ distributions and FRB energy functions. Our estimate for the lower limit on $E^{\mathrm{max}}_{\mathrm{iso}}$ ranges between $1.2\times10^{41} - 1.9\times10^{42}$\,erg. This is better visualized in Figure~\ref{fig:ll_vs_out} which shows how the lower limit on $E^{\mathrm{max}}_{\mathrm{iso}}$ varies as a function of the number of outliers removed. Our best estimate for the lower-limit on $E^{\mathrm{max}}_{\mathrm{iso}}$ is $1.2\times10^{42}$\,erg, obtained after removing the number of outliers informed from the simulated sample generated using the best-fit $\text{DM}_{\text{host}}$ and energy distributions from \cite{Shin_FRB_pop}.   

For the 17 simulated samples, the number of outliers range between 0 and 1234. The outlier removal method ensures that the lower limit on $E^{\mathrm{max}}_{\mathrm{iso}}$ for these simulated samples is consistent with the true value in the sample. From Table~\ref{tab:Emax_simulated}, it is clear that this lower-limit is dependent on the assumed FRB luminosity function. Higher $E^{\mathrm{cut-off}}_{\mathrm{iso}}$ and $\gamma$ give simulated FRBs with higher energies leading to higher values for the lower limit on $E^{\mathrm{max}}_{\mathrm{iso}}$. Moreover, the number of outliers is also sensitive to the assumed FRB luminosity function, such that simulations with $\gamma = 0.99$ have the highest number of outliers. Such a $\gamma$ allows for more FRBs to have energies close to the true $E^{\mathrm{max}}_{\mathrm{iso}}$ in the sample, such that only a small excess DM from the host or intervening halos is required to push their inferred energies above the true $E^{\mathrm{max}}_{\mathrm{iso}}$. On the contrary, the simulations with $\gamma = -2.04$ have the least number of outliers because there are very few FRBs with energies close to the true $E^{\mathrm{max}}_{\mathrm{iso}}$. Our results show that the number of outliers is also sensitive to the assumed $\mathrm{DM}_{\mathrm{host}}$ distribution, such that simulations with larger values of $\mu_{\mathrm{host}}$ and $\sigma_{\mathrm{host}}$ have a larger number of outliers. 

An important caveat in applying the framework described in Section~\ref{framework} on Catalog 2 FRBs is that the fluences in Catalog 2 are lower limits calculated under the assumption that the FRBs are detected at the location of the highest sensitivity within CHIME's primary beam \citep{bridget_flux}. A subset of these FRBs have baseband voltage data and thus more precise localizations, allowing the use of the accurate location of FRBs within the primary beam to estimate fluence values rather than lower limits \citep{basecat1}. The increase factor between the baseband fluences and the lower limits on fluence is dependent on where within the primary beam the FRBs are detected, and ranges between 1 and $\sim$300, with only $\sim15\%$ of bursts having increase factor $>10$. We estimate the lower limit on $E^{\mathrm{max}}_{\mathrm{iso}}$ for Catalog 2 FRBs using the lower limits on fluences and then multiplying the resulting value by the median increase factor of $3$. 

\begin{figure}[t]
\centering
\includegraphics[width=1\textwidth]{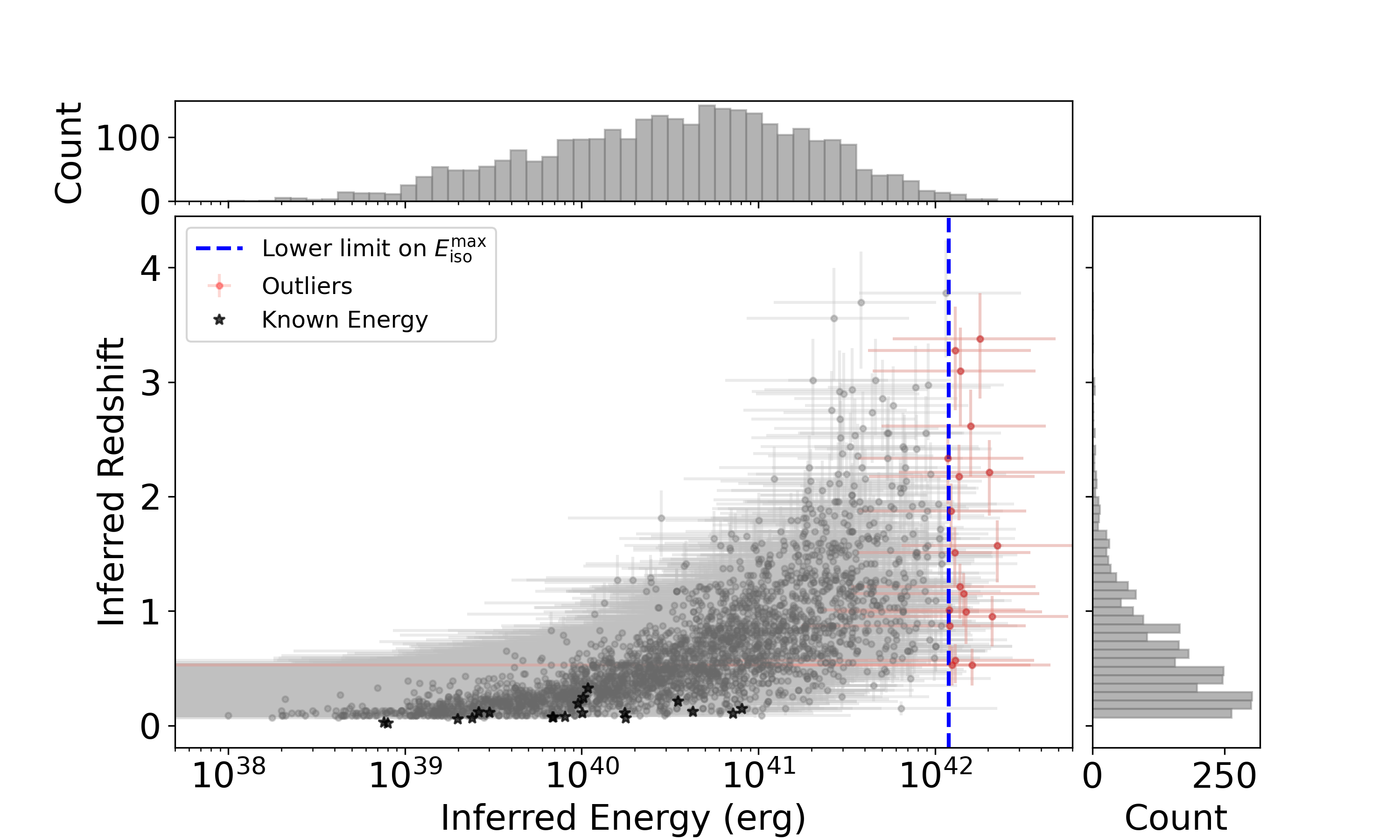}
\caption{For the 2,998 Catalog 2 FRBs, the median $z$ and energy from $ P(z \mid \mathrm{DM})$ and $ P(E \mid \mathrm{DM})$, respectively, are plotted, along with the 68\% confidence intervals. The red points show the 19 outliers that were removed for our best estimate of the lower limit on $E^{\mathrm{max}}_{\mathrm{iso}}$ of $1.2\times10^{42}$\,erg (see~Table \ref{tab:Emax_simulated}), which is shown by the vertical blue dashed line. The black stars mark the 19 FRBs that have confident hosts with known $z$ and robust energy measurements, enabled by localisations with the CHIME--KKO baseline. Irrespective of the inferred $z$, the inferred energies of the FRBs appear collectively limited around $\sim10^{42}$\,erg. \label{fig:z_E}}
\end{figure}

In Figure~\ref{fig:z_E}, we show the median $z$ and energy for the Catalog 2 FRBs from $P(z \mid \mathrm{DM})$ and $P(E_\text{iso} \mid \mathrm{DM})$, respectively, along with their 68\% confidence intervals taking into account the 68\% confidence interval for increase factor to estimate the error bars on energy. The vertical blue dashed line shows our best estimate for the lower limit on $E^{\mathrm{max}}_{\mathrm{iso}}$ of $1.2\times10^{42}$\,erg from Catalog 2 FRBs, after removing 19 outliers (red points). Additionally, the black stars show the $z$ and energies of 19 FRBs with confident host associations that were localized using interferometry between CHIME and the KKO Outrigger \citep{KKO_overview, KKO_catalog}. These local-Universe FRBs with relatively low $z$ agree with the low-energy end of the inferred redshift-energy distribution for Catalog 2 FRBs. An interesting observation in Figure~\ref{fig:z_E} is the turnover of the $z$-energy data points at higher energies such that irrespective of the inferred $z$, the inferred energies do not go far beyond $\sim10^{42}$\,erg, which is in agreement with our best estimate for the lower limit on $E^{\mathrm{max}}_{\mathrm{iso}}$. This suggests that we are already probing the maximum energy of FRBs using the available sample of FRBs. 

An important consideration in such an analysis is the selection biases introduced by the instrument. For example, \cite{Shin_FRB_pop} show that for CHIME/FRB, there is a ``turnover" in the DM-z relationship above DM $\sim$ 2000\,pc\,cm$^{-3}$ such that the Macquart relationship is no longer valid. Such FRBs are more likely to occur at $z < 2$, rather than at $z > 2$ as expected from the Macquart relation. The effect of this on our lower limit on $E^{\mathrm{max}}_{\mathrm{iso}}$ estimate from Catalog 2 FRBs is taken into account by removing outliers. The FRBs in the simulated sample which have $z$ inferred from $P(z \mid \mathrm{DM})$ larger than the true value, and which falsely inflate the lower limit on $E^{\mathrm{max}}_{\mathrm{iso}}$ value are considered outliers. Our method estimates the number of such outliers, which are then removed from the Catalog 2 sample before estimating the lower limit on $E^{\mathrm{max}}_{\mathrm{iso}}$. Another way to deal with this selection effect could be removing all FRBs having DM $>$ 2000\,pc\,cm$^{-3}$ from the Catalog 2 sample before obtaining the lower limit on $E^{\mathrm{max}}_{\mathrm{iso}}$. Doing so, we obtain a lower limit on $E^{\mathrm{max}}_{\mathrm{iso}}$ of $1.7\times10^{42}$\,erg. This is consistent with the higher-end of the range of values we obtained in Table~\ref{tab:Emax_simulated}. This is because removing FRBs with DM $>$ 2000\,pc\,cm$^{-3}$ would exclude the FRBs with the highest inferred $z$. However, as seen from Figure~\ref{fig:z_E}, there are FRBs with relatively lower inferred $z$ which have the highest inferred energies. Thus, our outlier removal method presents a more conservative range of values for the lower limit on $E^{\mathrm{max}}_{\mathrm{iso}}$. 

\begin{figure}[t]
    \centering
    \subfigure[]{\includegraphics[width=0.45\textwidth]{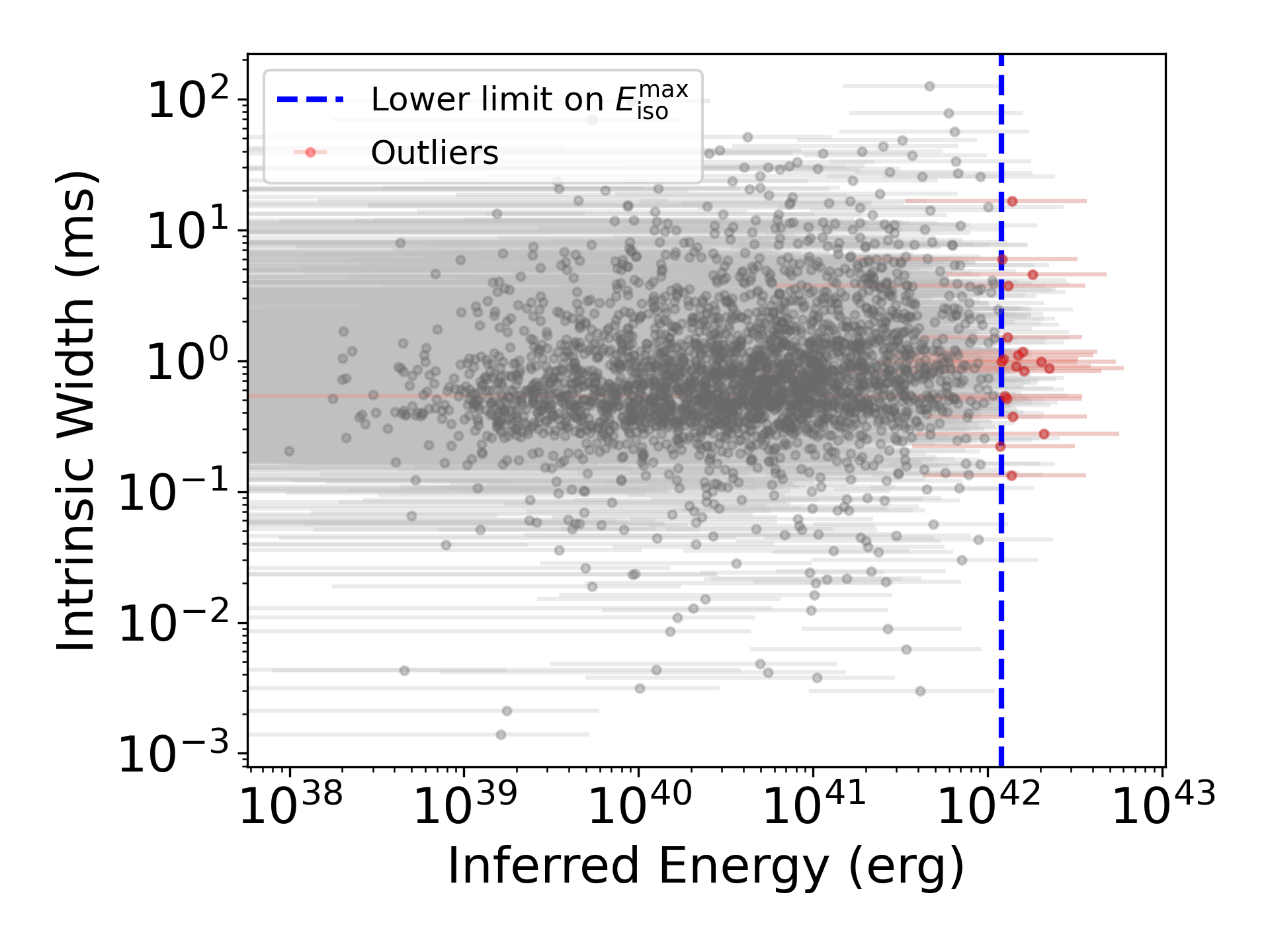}\label{fig:E_vs_intr_width}} 
    \subfigure[]{\includegraphics[width=0.45\textwidth]{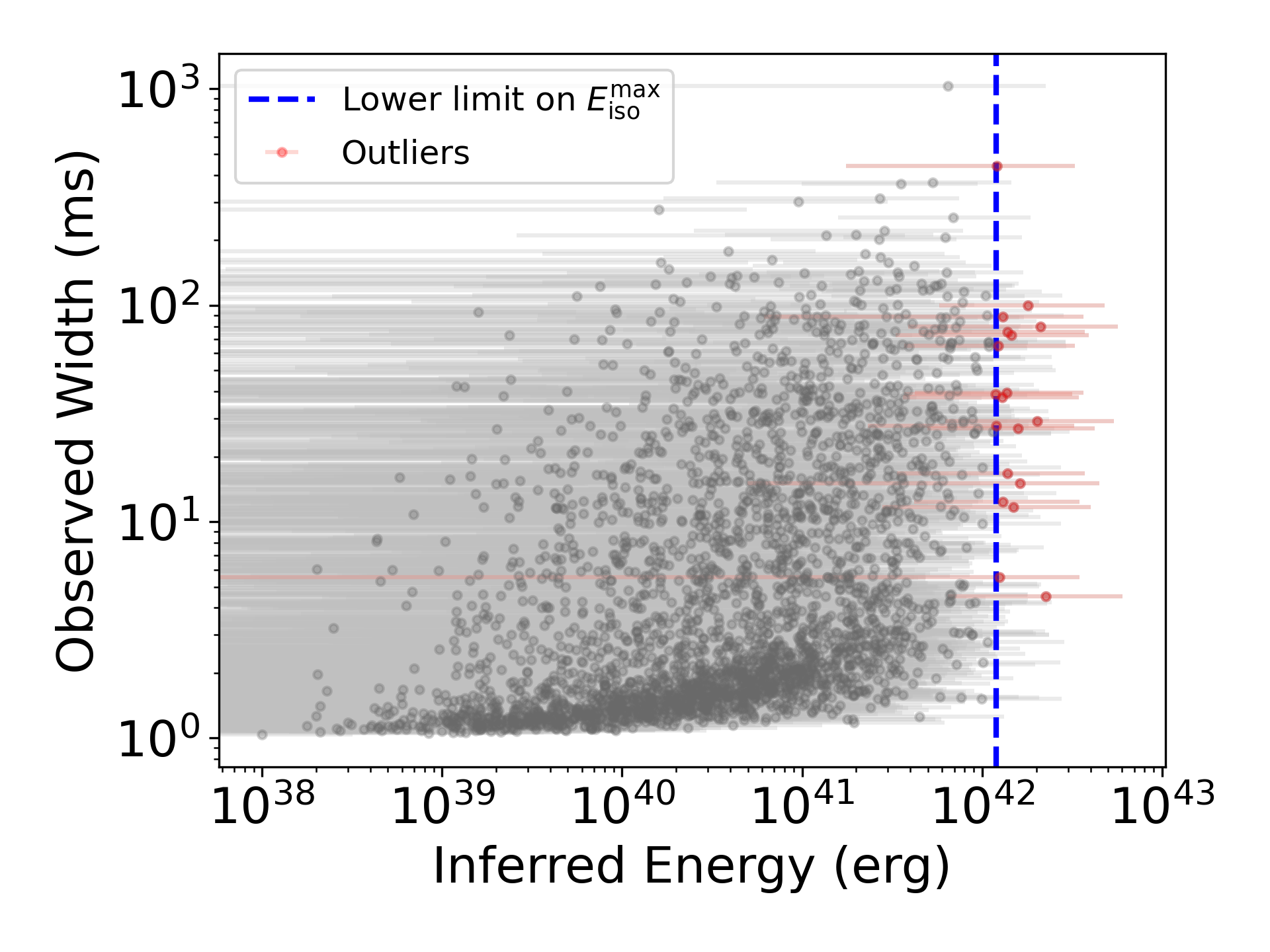}\label{fig:E_vs_obs_width}} 
    \caption{The median inferred energies with 68\% confidence interval as a function of the intrinsic widths (left panel) and observed widths (right panel) of Catalog 2 FRBs. The red points mark the 19 outliers that were removed to obtain our best estimate for the lower limit on $E^{\mathrm{max}}_{\mathrm{iso}}$ (blue dashed line). There appears to be no strong correlation between the energies and widths of the highest inferred energy FRBs. \label{fig:E_vs_width}} 
\end{figure}

Moreover, to ensure that we are not missing out on the highest energy bursts due to CHIME/FRB's selection bias against wider bursts, we plot the median inferred energies with 68\% confidence interval as a function of the intrinsic widths and observed widths of Catalog 2 FRBs in Figure~\ref{fig:E_vs_width}. \cite{chime_cat2} presents the intrinsic width estimates for the FRBs and we obtain the observed widths by adding the intrinsic widths, intrachannel DM smearing, scattering timescale, and the time resolution of the data in quadrature \citep{emmanuel_fitburst}. In Figure \ref{fig:E_vs_intr_width}, it is clear that there is no strong correlation between the intrinsic widths and inferred energies of the FRBs, with the Pearson correlation coefficient\footnote{A Pearson correlation coefficient evaluates the strength of a linear relationship between two variables. Its value can range between -1 and 1 with -1 indicating a perfect negative correlation and +1 indicating a perfect positive correlation. Values close to 0 indicate no correlation.} being 0.08.  In Figure \ref{fig:E_vs_obs_width}, it might appear that there is a positive correlation between the observed widths and inferred energies due to the lack of high-width and low-energy data points. The Pearson correlation coefficient in this case is 0.27. The deficit of data points at the low-energy end is an observational bias due to the difficulty in detecting faint bursts that are wider. On the other hand, bursts with the highest inferred energies span the entire range of observed width values. To eliminate the bias at the low-energy end, we calculate the Pearson correlation coefficient for bursts with inferred energies greater than $10^{40}$\,erg, and obtain a value of 0.09. Thus, our results do not seem affected by the detection biases of CHIME/FRB. 

\begin{deluxetable}{ c c c c c c c }
\tabletypesize{\small}
\tablewidth{0pt} 
\tablecaption{Results of the framework applied on simulated FRB samples and the Catalog 2 FRB sample.\label{tab:Emax_simulated}}
\tablehead{
\colhead{ log$_{10}$($\mu_{\mathrm{host}}$)} & \colhead{log$_{10}$($\sigma_{\mathrm{host}}$)} & \colhead{ log$_{10}$($E^{\mathrm{cut-off}}_{\mathrm{iso}}$)} & \colhead{$\gamma$} & \colhead{No. of outliers} & \colhead{Simulated sample:} & \colhead{Catalog 2:} \\
\colhead{} & \colhead{} & \colhead{} & \colhead{} & \colhead{} & \colhead{log$_{10}$(Lower limit on $E^{\mathrm{max}}_{\mathrm{iso}}$)} &  \colhead{log$_{10}$(Lower limit on $E^{\mathrm{max}}_{\mathrm{iso}}$)} \\
\colhead{} & \colhead{} & \colhead{(erg)} & \colhead{} & \colhead{} & \colhead{(erg)} & \colhead{(erg)}}
\startdata
\textbf{1.93} & \textbf{0.41} & \textbf{41.38} & \textbf{-1.30} & \textbf{19} & \textbf{42.07} & \textbf{42.09} \\
1.34 & 0.08 & 40.40 & -2.04 & 76 & 40.91 & 41.92 \\
1.34 & 0.08 & 40.40 & 0.99 & 159 & 41.48 & 41.74 \\
1.34 & 0.08 & 42.46 & -2.04 & 0 & 42.63 & 42.28 \\
1.34 & 0.08 & 42.46 & 0.99 & 10 & 43.48 & 42.15 \\
1.34 & 0.82 & 40.40 & -2.04 & 136 & 41.08 & 41.77 \\
1.34 & 0.82 & 40.40 & 0.99 & 452 & 41.47 & 41.53 \\
1.34 & 0.82 & 42.46 & -2.04 & 1 & 42.65 & 42.26 \\
1.34 & 0.82 & 42.46 & 0.99 & 12 & 43.54 & 42.13 \\
2.40 & 0.08 & 40.40 & -2.04 & 234 & 41.04 & 41.68 \\
2.40 & 0.08 & 40.40 & 0.99 & 1117 & 41.49 & 41.16 \\
2.40 & 0.08 & 42.46 & -2.04 & 3 & 42.66 & 42.22 \\
2.40 & 0.08 & 42.46 & 0.99 & 54 & 43.45 & 41.97 \\
2.40 & 0.82 & 40.40 & -2.04 & 625 & 40.93 & 41.43 \\
2.40 & 0.82 & 40.40 & 0.99 & 1234 & 41.38 & 41.10 \\
2.40 & 0.82 & 42.46 & -2.04 & 0 & 42.61 & 42.28 \\
2.40 & 0.82 & 42.46 & 0.99 & 35 & 43.53 & 42.01 \\
\enddata
 % \hline
\tablecomments{$\mu_{\mathrm{host}}$ and $\sigma_{\mathrm{host}}$ describe the mean and standard deviation for the log-normal distribution for $\text{DM}_{\text{host}}$, $E^{\mathrm{cut-off}}_{\mathrm{iso}}$ describes the cut-off energy, and $\gamma$ describes the power-law index of the Schechter function used in each simulation. The lower limit on $E^{\mathrm{max}}_{\mathrm{iso}}$ for the simulated samples and Catalog 2 FRBs is obtained after removing the corresponding number of outliers. For Catalog 2 FRBs, the lower limit on $E^{\mathrm{max}}_{\mathrm{iso}}$ includes the median increase factor between the baseband fluences and lower limits on fluences. The simulated sample obtained using the best-fit parameter values from \cite{Shin_FRB_pop} and the resulting best estimate for the lower-limit on $E^{\mathrm{max}}_{\mathrm{iso}}$ are marked in bold.}
\end{deluxetable}

\section{Discussion}\label{Discussion}

The maximum energy of FRBs can be an important indicator of the energy reservoir of FRB sources and thus provides crucial insights into their origins. We therefore compare our best estimate for the lower limit on $E^{\mathrm{max}}_{\mathrm{iso}}$ of $\sim10^{42}$\,erg with other works studying FRB energetics and discuss the resulting implications for FRB progenitors and emission mechanisms. 

\subsection{Comparison to Other Works}\label{comparision}
Various works have studied the energetics of FRBs by modeling the FRB energy distribution with a Schechter function. The motivation to use this functional form is to allow for FRBs with energies higher than $E^{\mathrm{cut-off}}_{\mathrm{iso}}$, but with orders of magnitude lower rates. Similar to the lower limit on $E^{\mathrm{max}}_{\mathrm{iso}}$, $E^{\mathrm{cut-off}}_{\mathrm{iso}}$ is not equal to but close to $E^{\mathrm{max}}_{\mathrm{iso}}$. These other works use FRBs detected at different radio frequencies and with different observing bandwidths. Under our previously stated assumptions of $\alpha = 0$ and $\Delta\nu = 1$\,GHz, we compare $E^{\mathrm{cut-off}}_{\mathrm{iso}}$ obtained in these works with our lower limit on $E^{\mathrm{max}}_{\mathrm{iso}}$. 

\citet{Luo_2020} used 46 FRBs detected with various instruments, including the upgraded Molonglo Observatory Synthesis Telescope in the Southern-Sky survey \citep[UTMOST-SS,][]{Caleb_2017} and Australian Square Kilometre Array Pathfinder in the Commensal Real-time ASKAP Fast Transient survey \citep[ASKAP/CRAFT,][]{Shannon_2018}, taking into account their selection biases and found log$_{10}$\,$E^{\mathrm{cut-off}}_{\mathrm{iso}} = 42.08\substack{+0.3 \\ -0.06}$\,erg. \citet{James_2022} used 16 ASKAP/CRAFT FRBs localized to their host galaxies and 60 poorly localized FRBs with no known hosts from Parkes and ASKAP/CRAFT and found log$_{10}$\,$E^{\mathrm{cut-off}}_{\mathrm{iso}} = 41.26\substack{+0.27 \\ -0.22}$\,erg. \citet{Shin_FRB_pop} used 225 FRBs presented in the First CHIME/FRB Catalog \citep{cat1} calibrated for selection effects and found log$_{10}$\,$E^{\mathrm{cut-off}}_{\mathrm{iso}} = 41.38\substack{+0.51 \\ -0.51}$\,erg. While these works have studied the energy function of a population of one-off FRBs, another approach is to use bursts from individual repeating FRB sources. \citet{Omar_2024} observed the repeating FRB\,20220912A for 1,500 hours and detected 130 high-energy bursts having log$_{10}$\,$E^{\mathrm{cut-off}}_{\mathrm{iso}}$ of 41.32$\substack{+0.45 \\ -0.30}$\,erg. It is remarkable that using a $>10\times$ larger sample than previous studies, the range of log$_{10}$\,(lower limit on $E^{\mathrm{max}}_{\mathrm{iso}}$) obtained in this work and presented in Table~\ref{tab:Emax_simulated} is consistent within error bars with the log$_{10}$\,$E^{\mathrm{cut-off}}_{\mathrm{iso}}$ obtained in other works. 
 
In addition to these FRB population studies, there have also been studies where individual high-energy bursts from one-off and repeating FRBs have been reported\footnote{For works in which spectral energy has been reported, we assume an intrinsic bandwidth of 1\,GHz for the FRB to obtain their energies.}. \citet{Ravi_2019} reported a burst detected and localized using the Deep Synoptic Array (DSA) having log$_{10}$\,$E_{\mathrm{iso}}$ = 42.75\,erg. \citet{Ryder_2023} present an FRB originating at $z=1$ detected by ASKAP/CRAFT with log$_{10}$\,$E_{\mathrm{iso}}$ = 42.30\,erg. In terms of repeating FRBs, in the sample of high-energy bursts from FRB\,20201124A presented in \citet{Kirsten_2024}, the highest-energy burst has log$_{10}$\,$E_{\mathrm{iso}}$ = 41.49\,erg. The log$_{10}$\,(lower limit on $E^{\mathrm{max}}_{\mathrm{iso}}$) obtained from Catalog 2 FRBs is thus similar to the energies of highly energetic one-off and repeating FRB bursts.

\subsection{Implications for the Emission Mechanisms and Progenitors of FRBs}

In this work, we study the largest sample of one-off FRBs containing 2,998 unique sources  to constrain $E^{\mathrm{max}}_{\mathrm{iso}}$. It is informative that the range of lower limit on $E^{\mathrm{max}}_{\mathrm{iso}}$ that we obtain in this work is consistent with the various studies of FRB energetics done using FRBs detected with different instruments with their own intrinsic selection effects. This suggests there is a limit to the amount of energy that FRB sources can produce. This is also apparent in Figure~\ref{fig:z_E}, where the inferred energies do not go far beyond $\sim10^{42}$\,erg even for FRBs with high inferred $z$, providing further indication of the energy reservoir of FRB sources.   

Assuming the radio efficiency factor for FRB sources to be $\eta_{\mathrm{radio}} \lesssim 10^{-4}$ based on the ratio of energies released in the radio and X-ray bursts from the Galactic magnetar SGR\,1935+2154 \citep{Bochenek_SGR}, and adopting similar beaming fractions for radio and high-energy emission, this gives an isotropic equivalent source energy of $E^{\mathrm{src}}_{\mathrm{iso}} \gtrsim 10^{46}$\,erg. The total external dipolar magnetic energy of a magnetar is estimated to be $E_{dip} = 1/6B_P^2R^3 = (1.7\times10^{47}\,\mathrm{erg})B_{P,15}^2R_6^3$ for a surface dipole magnetic field strength of $B_P = 10^{15}$\,G and a neutron star radius of $R = 10^6$\,cm. It is also possible for a magnetar to have local regions of higher magnetic field strengths from multipolar field components \citep{Vicky_magnetar}. Thus, magnetars have the required energy reservoir to power such energetic FRBs. This is further proven by the fact that the most energetic magnetar giant flares ever observed have isotropic equivalent energies ranging over $10^{46}-10^{47}$\,erg \citep{Hurley_giant_flare, Ofek_MGF}. The most energetic FRB bursts could thus be produced via forward shocks when these extremely energetic magnetar giant flares collide with the surrounding medium \citep{Metzger_2019}. On the other hand, \cite{KUmar_2017} estimate a beaming-corrected energy release of $\sim10^{36}$\,erg in emission mechanisms involving reconnection of magnetic fields in the magnetosphere of magnetars, possibly suggesting different magnetar emission mechanisms for lower and higher energy FRBs. 

Another interesting observation from Section~\ref{comparision} is that one-off FRBs and bursts from repeating FRB sources have similar maximum energies. This could imply that one-off and repeating FRB sources have the same energy reservoir, which may suggest a common progenitor, but possibly different emission mechanisms, given observed differences in burst morphologies of one-off and repeat bursts \citep{Ziggy_repeater_morphology, Curtin_2025}. A common progenitor scenario is also consistent with the conclusions from \citet{Kirsten_2024} and \citet{Omar_2024} that one-off FRBs could be the highest energy bursts from repeating sources that appear as one-offs due to their lower rates.

The energy reservoir of FRB sources is also revealed by the sum total of energy of repeat bursts. \citet{FAST_R147} observed 11,553 bursts from the hyper-active repeating source FRB\,20240114A with isotropic energies ranging over $10^{37}-10^{40}$\,erg, assuming an intrinsic burst bandwidth of 1\,GHz. The total energy of all bursts combined is $2.5\times10^{42}$\,erg, similar to that of the most energetic one-off FRBs. FRB sources can thus produce thousands of low-energy bursts releasing the same amount of energy as a single high-energy burst, similar to hundreds of X-ray bursts observed during magnetar burst storms producing the same amount of energy as a single very energetic flare \citep{Younes_SGR_storm, Hu_SGR_flare}. On the other hand, \citet{Omar_2024} observed 130 high-energy bursts from FRB\,20220912A having total isotropic energy of $1.3\times10^{43}$\,erg. An explanation for such high energies could be that the magnetic field powering the repeat bursts is replenished over the activity period of the repeater. If a one-off FRB with energy $10^{43}-10^{44}$\,erg is observed where replenishment cannot occur, it could imply unusually high $\eta_{\mathrm{radio}}$ and/or $B_P$, or extremely beamed emission such that $E^{\mathrm{src}}$ is still within what a typical magnetar could produce \citep{NS_beaming}. Thus, such highly energetic bursts, if detected, would provide stringent constraints on the parameters in FRB emission mechanism models involving magnetar progenitors. Alternatively, such bursts, whose non-detection points towards a very low rate of occurrence, could possibly originate from a different progenitor altogether. 

\subsection{FRB Energetics in the Era of CHIME/FRB Outriggers}

The accurate estimation of FRB energies requires their $z$ to be known. While in this work we combined the DMs and fluences of 2,998 FRBs to estimate a lower bound on $E^{\mathrm{max}}_{\mathrm{iso}}$, knowing the $z$ and thus the energies for such a large sample of FRBs would either confirm that there is a limit of $\sim10^{42}$\,erg for the most energetic FRBs, or reveal the presence of even more energetic bursts. CHIME/FRB's FRB detection rate of $\sim3$ FRBs per day and its fluence threshold of $\sim3.5$\,Jy\,ms means that it detects many FRBs at relatively higher energies compared to other telescopes. The CHIME/FRB Outriggers, which are smaller CHIME-like telescopes forming a very-long-baseline interferometry network with CHIME, provide a FRB localization accuracy of $\sim$50\,mas $\times$ 100\,mas \citep{Outriggers_design_overview} and will enable a $z$ determination for many of these FRBs. Such a precise localization can also identify the local environments of FRBs within their host galaxies \citep{RBFLOAT}. This opens up the possibility of studying not just the burst properties, but also the locations of highly energetic FRBs, shedding more light on their progenitors and emission mechanisms. A distribution of FRB energies and their local environments could be useful to identify the existence of progenitor(s) that create extremely rare but much more energetic bursts than typical FRBs. The era of CHIME/FRB Outriggers is thus going to be game-changing for identifying the most energetic FRBs and using them to understand the origins of FRBs. 

\section{Summary and Conclusion}

The Second CHIME/FRB Catalog presents the largest currently available sample of FRBs along with their burst morphology parameters including DMs and lower limits on fluence. We use 2,998 one-off FRBs from this sample and develop a framework to combine their DMs and fluences with $ P(z \mid \mathrm{DM})$ to obtain a lower limit on the maximum energy of FRBs. In order to verify the robustness of this framework, we generate 17 simulated FRB samples assuming different parameter values in the Schechter function for the FRB energy function and the log-normal distribution for $\mathrm{DM}_{\mathrm{host}}$. Applying our framework on these simulated samples allows us to identify the number of outliers, which are FRBs that have large DM contributions from the host galaxy and/or intervening galaxy halos, which when removed, give us a lower limit on $E^{\mathrm{max}}_{\mathrm{iso}}$ consistent with the true $E^{\mathrm{max}}_{\mathrm{iso}}$ in the sample. The lower limit on $E^{\mathrm{max}}_{\mathrm{iso}}$ for the Catalog 2 sample, obtained after removing the number of outliers corresponding to the 17 simulated samples, ranges between $1.2\times10^{41}$ and $1.9\times10^{42}$\,erg, with our best estimate being $1.2\times10^{42}$\,erg. Moreover, inferring the $z$ and $E_\text{iso}$ from $P(z \mid \mathrm{DM})$ and $P(E_\text{iso} \mid \mathrm{DM})$, respectively, reveals that irrespective of the inferred $z$, the inferred energies of Catalog 2 FRBs are collectively limited around $\sim10^{42}$\,erg. 

Our constraint on $E^{\mathrm{max}}_{\mathrm{iso}}$ is consistent with those obtained from other works involving much smaller populations of one-off FRBs, repeat bursts from repeating FRBs, and individual highly energetic FRBs. This agreement between our work, based on by far the largest available statistical sample, and various studies of FRB energetics involving FRBs detected with different instruments with their own intrinsic selection effects, might imply that current observations are able to probe the maximum energy achievable by FRBs. Moreover, similar maximum energy estimates between one-off and repeating FRBs might imply a common progenitor for these two populations of FRBs. It is also interesting that the amount of energy released by hyper-active repeating FRBs in hundreds-to-thousands of bursts can be released in an individual highly energetic burst. Under standard assumptions of radio efficiencies and dipolar magnetic fields, the most energetic FRBs are consistent with being produced from the most energetic magnetar giant flares; however, it is also reaching the limits of the amount of energy magnetars could produce. Detecting bursts with energies an order of magnitude or more higher could imply extreme scenarios such as an unusually large dipolar magnetic field, very small beaming fraction, or energy released from multi-pole field components for magnetar progenitors. It is also possible that such bursts, whose non-detection till now points towards a very low occurrence rate, come from a different progenitor altogether. CHIME/FRB and its Outriggers, with their excellent FRB detection rate and localization capability, will be instrumental in detecting the highest energy FRBs and constraining their progenitor models and emission mechanisms. 

\vspace{1cm}
\section*{Acknowledgments}
We acknowledge that CHIME is located on the traditional, ancestral, and unceded territory of the Syilx/Okanagan people. We are grateful to the staff of the Dominion Radio Astrophysical Observatory, which is operated by the National Research Council of Canada. CHIME operations are funded by a grant from the NSERC Alliance Program and by support from McGill University, University of British Columbia, and University of Toronto. CHIME was funded by a grant from the Canada Foundation for Innovation (CFI) 2012 Leading Edge Fund (Project 31170) and by contributions from the provinces of British Columbia, Québec and Ontario. The CHIME/FRB Project was funded by a grant from the CFI 2015 Innovation Fund (Project 33213) and by contributions from the provinces of British Columbia and Québec, and by the Dunlap Institute for Astronomy and Astrophysics at the University of Toronto. Additional support was provided by the Canadian Institute for Advanced Research (CIFAR), the Trottier Space Institute at McGill University, and the University of British Columbia. The CHIME/FRB baseband recording system is funded in part by a CFI John R. Evans Leaders Fund award to IHS.

The AstroFlash research group at McGill University, University of Amsterdam, ASTRON, and JIVE is supported by: a Canada Excellence Research Chair in Transient Astrophysics (CERC-2022-00009); an Advanced Grant from the European Research Council (ERC) under the European Union’s Horizon 2020 research and innovation programme (`EuroFlash'; Grant agreement No. 101098079); and an NWO-Vici grant (`AstroFlash'; VI.C.192.045).

V.S. is supported by a Fonds de Recherche du Quebec - Nature et Technologies (FRQNT) Doctoral Research Award. V.M.K. holds the Lorne Trottier Chair in Astrophysics \& Cosmology, a Distinguished James McGill Professorship, and receives support from an NSERC Discovery grant (RGPIN 228738-13). K.W.M. is supported by NSF Grant Nos. 2008031, 2510771 and holds the Adam J. Burgasser Chair in Astrophysics. M.W.S is a Fonds de Recherche du Quebec - Nature et Technologies (FRQNT) postdoctoral fellow and acknowledges support from the Trottier Space Institute Fellowship program. A.P.C. is a Canadian SKA Scientist and is funded by the Government of Canada / est financé par le gouvernement du Canada. K.T.M is supported by a FRQNT Master’s Research Scholarship. M.N. is a Fonds de Recherche du Quebec - Nature et Technologies (FRQNT) postdoctoral fellow. K.N. acknowledges support by NASA through the NASA Hubble Fellowship grant \# HST-HF2-51582.001-A awarded by the Space Telescope Science Institute, which is operated by the Association of Universities for Research in Astronomy, Incorporated, under NASA contract NAS5-26555. A.P. is a Trottier Space Institute Postdoctoral Fellow. A.B.P. acknowledges support by NASA through the NASA Hubble Fellowship grant HST-HF2-51584.001-A awarded by the Space Telescope Science Institute, which is operated by the Association of Universities for Research in Astronomy, Inc., under NASA contract NAS5-26555. A.B.P. also acknowledges prior support from a Banting Fellowship, a McGill Space Institute~(MSI) Fellowship, and a Fonds de Recherche du Quebec -- Nature et Technologies~(FRQNT) Postdoctoral Fellowship.

\appendix
\section{Derivation of $P(\mathrm{DM} \mid z)$} 
\label{sec:P_DM_z}
Marginalizing over $\mathrm{DM}_{\mathrm{MW}}$, $\mathrm{DM}_{\mathrm{cosmic}}$, and $\mathrm{DM}_{\mathrm{host}}$ and using the rules of conditional probability:
\begin{equation}\label{eq:P_DM_z}
\begin{split}
    P(\mathrm{DM}|z) &= \int\int\int d\mathrm{DM}_{\mathrm{MW}}\,d\mathrm{DM}_{\mathrm{cosmic}}\,d\mathrm{DM}_{\mathrm{host}}\,P(\mathrm{DM}, \mathrm{DM}_{\mathrm{MW}}, \mathrm{DM}_{\mathrm{cosmic}}, \mathrm{DM}_{\mathrm{host}} \mid z) \\
    &\propto \int\int\int d\mathrm{DM}_{\mathrm{MW}}\,d\mathrm{DM}_{\mathrm{cosmic}}\,d\mathrm{DM}_{\mathrm{host}}\,P(\mathrm{DM} \mid \mathrm{DM}_{\mathrm{MW}}, \mathrm{DM}_{\mathrm{cosmic}}, \mathrm{DM}_{\mathrm{host}}, z)\times \\
    &\quad P(\mathrm{DM}_{\mathrm{MW}} \mid \mathrm{DM}_{\mathrm{cosmic}}, \mathrm{DM}_{\mathrm{host}}, z)\,P(\mathrm{DM}_{\mathrm{cosmic}} \mid \mathrm{DM}_{\mathrm{host}}, z)\, P(\mathrm{DM}_{\mathrm{host}} \mid z) \\
\end{split}
\end{equation}
Since the total DM of the FRB has to satisfy Equation~\ref{total_DM}, $P(\mathrm{DM} \mid \mathrm{DM}_{\mathrm{MW}}, \mathrm{DM}_{\mathrm{cosmic}}, \mathrm{DM}_{\mathrm{host}}, z)$ becomes a delta function. $\mathrm{DM}_{\mathrm{cosmic}}$ is dependent on $z$ but not on any other DM components. $\mathrm{DM}_{\mathrm{MW}}$ and $\mathrm{DM}_{\mathrm{host}}$ are independent of other DM components, as well as $z$. Taking this into account, Equation~\ref{eq:P_DM_z} becomes
\begin{equation}
\begin{split}
P(\mathrm{DM}|z) &\propto \int\int d\mathrm{DM}_{\mathrm{MW}}\,d\mathrm{DM}_{\mathrm{cosmic}}\,d\mathrm{DM}_{\mathrm{host}}\, \delta\left(\mathrm{DM} - \mathrm{DM}_{\mathrm{cosmic}} - \mathrm{DM}_{\mathrm{MW}} - \frac{\mathrm{DM}_{\mathrm{host}}}{1+z}\right)\,P( \mathrm{DM}_{\mathrm{MW}})\times \\ &\quad P( \mathrm{DM}_{\mathrm{cosmic}} \mid z)\,P( \mathrm{DM}_{\mathrm{host}}) 
\end{split}
\end{equation}
Integrating over $\mathrm{DM}_{\mathrm{cosmic}}$ resolves the delta function and gives
\begin{equation}
    P(\mathrm{DM}|z) \propto \int \int d\mathrm{DM}_{\mathrm{MW}}\,d\mathrm{DM}_{\mathrm{host}}\,P( \mathrm{DM}_{\mathrm{MW}})\,P\left( \mathrm{DM}_{\mathrm{cosmic}} = \mathrm{DM} - \mathrm{DM}_{\mathrm{MW}} - \frac{\mathrm{DM}_{\mathrm{host}}}{1+z} \mid z\right)\,P( \mathrm{DM}_{\mathrm{host}})
\end{equation}

\section{Derivation of $P(E_\text{iso} \mid \mathrm{DM})$}\label{sec:P_E_DM}
We start by using Equation~\ref{E} to derive $\left| \frac{dE_\text{iso}}{dz}\right|$:
\begin{equation}\label{dE_dz}
    \frac{dE_\text{iso}}{dz} = 4\pi F_{\nu}\Delta\nu\left(\frac{2\text{D}_{\text{L}}}{(1+z)^2}\frac{d\text{D}_{\text{L}}}{dz} - \frac{2\text{D}_{\text{L}}^2}{(1+z)^3} \right)
\end{equation}
Under standard cosmology assumptions, the luminosity distance $\text{D}_{\text{L}}$ depends on $z$ as:
\begin{equation}\label{D_L}
    \text{D}_{\text{L}}(z) = (1+z)\int^z_0 \frac{c\,dz}{H_0\sqrt{\Omega_m(1+z)^3 + \Omega_{\Lambda}}}
\end{equation}
such that
\begin{equation}\label{dD_L_dz}
    \frac{d\text{D}_{\text{L}}}{dz} = \frac{\text{D}_{\text{L}}}{1+z} + \frac{(1+z)c}{H_0\sqrt{\Omega_m(1+z)^3 + \Omega_{\Lambda}}}
\end{equation}
Combining Equations~\ref{dE_dz} and \ref{dD_L_dz} and substituting the $4\pi F_{\nu}\Delta\nu$ factor in terms of $E_\text{iso}$, we get:
\begin{equation}\label{dzdE}
    \frac{dz}{dE_\text{iso}} = \frac{1}{\frac{dE_\text{iso}}{dz}} = \frac{\text{D}_{\text{L}}H_0\sqrt{\Omega_m(1+z)^3 + \Omega_{\Lambda}}}{2E_\text{iso}(1+z)c}
\end{equation}
Substituting Equation~\ref{dzdE} in Equation~\ref{P_E_DM} gives us  $P(E_\text{iso} \mid \mathrm{DM})$. 

\bibliography{citations}{}
\bibliographystyle{aasjournal}

\end{document}